\documentclass{aastex701}
\newcommand{\superfit}{{\it Superfit}}
\usepackage{float}

\usepackage{tikz}
\usetikzlibrary{arrows.meta,positioning,shapes.geometric}

\begin{document}

\title{Superfit: A Supernova Spectral Classification Tool with Host-Galaxy Decomposition}
\shorttitle{Superfit}
\shortauthors{Howell et al.}

\author[orcid=0000-0003-4253-656X,sname='Howell']{D.~Andrew~Howell}
\affiliation{Las Cumbres Observatory, 6740 Cortona Dr. Suite 102, Goleta, CA}
\affiliation{Physics Department, University of California, Santa Barbara, Santa Barbara, CA}
\email[show]{ahowell@lco.global}

\author[orcid=0009-0001-8687-7764,sname='Kosobokova']{Alyona~Kosobokova}
\affiliation{Las Cumbres Observatory, 6740 Cortona Dr. Suite 102, Goleta, CA}
\affiliation{University of California, Santa Barbara, Santa Barbara, CA}
\email{alonakosobokova@gmail.com}

\author[orcid=0009-0004-9413-2801,sname='Goldwasser']{Samantha~Goldwasser}
\affiliation{Department of Particle Physics and Astrophysics, Weizmann Institute of Science
76100 Rehovot, Israel}
\email{goldwasser@coralcastlelabs.com}

\author[orcid=0000-0002-7996-8780,sname='Irani']{Ido~Irani}
\affiliation{Department of Particle Physics and Astrophysics, Weizmann Institute of Science
76100 Rehovot, Israel}
\email{idoirani@gmail.com}

\author[orcid=0009-0000-6389-1697,sname='Tiwary']{Saanvi~Tiwary}
\affiliation{Department of Statistics and Applied Probability, University of California, Santa Barbara,
Santa Barbara, CA, USA}
\email{sanvitiwary@gmail.com}

\author[orcid=0000-0002-0301-8017/,sname='Yaron']{Ofer Yaron}
\affiliation{Department of Particle Physics and Astrophysics, Weizmann Institute of Science
76100 Rehovot, Israel}
\email{ofer.yaron@weizmann.ac.il}

\author[orcid=0000-0002-3653-5598,sname='Gal-Yam']{Avishay Gal-Yam}
\affiliation{Department of Particle Physics and Astrophysics, Weizmann Institute of Science
76100 Rehovot, Israel}
\email{avishay.gal-yam@weizmann.ac.il}

\begin{abstract}
\superfit{} is a an open-source tool for automated spectroscopic classification of supernovae that accounts for redshift, reddening, and host-galaxy contamination.  Here we present the latest version, which has been rewritten in Python and is now available as a web-based application.  A user uploads an observed spectrum, and \superfit{} uses a curated library of transient and galaxy spectra to perform a simultaneous fit of transient and host galaxy spectra, while adjusting for color and redshift.  In contrast to machine-learning algorithms, \superfit{} uses a least squares fitting approach, which is reproducible, verifiable, and understandable to users.  The user may specify the binning of the spectra, set redshift and reddening limits, phase information, and restrict host galaxy and supernova templates for fitting.  A new, expanded template library includes modern transient subclasses and associated observational metadata. A modern browser-based interface allows users to configure fitting parameters, visualize spectra, monitor fitting progress, and interactively inspect and compare ranked candidate matches and their individual components. We describe the fitting methodology, template library, software architecture, and web interface, and demonstrate its capabilities through classification examples.
\end{abstract}


\section{Introduction}\label{sec:introduction}
For most of human history, when stellar deaths were bright enough to be seen by humans, they were seen as a mysterious, perhaps ominous new point of light in the sky. \citet{Tycho1572} termed these {\it stella novae}, meaning new stars, perhaps the most incorrectly named of all astrophysical phenomena.  It took until the early 1930s to realize that some of these events are extraordinarily more powerful than others, a new class called supernovae \citep{1934PNAS...20..254B}.  Further investigation revealed that some could be separated into multiple subclasses which turned out to be physically distinct \citep{1941PASP...53..224M}.  Since then a multitude of types have proliferated, almost always continuing the trend of a spectrosocopic type being named before it was physically understood.  

Since the features in a supernova spectrum vary with not just type, but phase, reddening, chemical abundances, stellar and environmental characteristics, temperature, and even instrumental and data reduction artifacts, they can be tricky to definitively categorize.  This was made all the more difficult with the routine discovery and observation of high redshift supernovae in the 1990s \citep[e.g.][]{1995ApJ...440L..41P}.  The light of distant host galaxies, angularly compressed into a smaller area, could easily swamp the light of a supernova, itself difficult to observe, even with the largest telescopes of the era.  Until this time, supernova types (and spectroscopic estimates of phases) were determined by eye by trained observers, and the process was something of an art form, requiring extensive observing and data reduction experience, an encyclopedic knowledge of supernovae, and a certain amount of gut instinct.  

To bring rigor and order to the process, and to more cleanly separate the supernova signal from the contaminating host galaxy spectrum, beginning around 2002, one of us (Howell) wrote the first supernova spectrum fitting code, \superfit{}.  The first version was written in IDL, complete with a graphical user interface and the ability to generate on-screen or PDF comparison plots.  Though it was initially written mainly to help the author classify SNe more rigorously, word spread of its utility, and in the era before online code repositories, the author personally distributed the code to those who requested it.  Eventually it was hosted at a website, and loosely described in \citet{Howell2005}.  

Initial development of \superfit\ experimented with cross-correlation, thought this was discarded because it required subtracting the continuum of a supernova, which can distort the spectrum, and discards key information.  Focusing on known features was also considered and rejected, because sometimes features are unknown, and when redshifts are unknown it can be difficult to map observed features to known supernova features.  The first public version of \superfit\ relied on using all information in an observed spectrum, fitting it to a linear combination of library supernova and host galaxy templates, while adjusting for reddening.  With this method, a user could gain greater intuition into whether the result output by the code was reliable -- they could see the observed spectrum  after deredshifting, dereddening, and host galaxy subtraction.  This was critical is judging the provenance of distant, noisy supernovae on the hairy edge of observability.

Other supernova fitting codes were soon developed, including {\it GELATO} \citep{2008A&A...488..383H}, and {\it SNID} \citet{2007ApJ...666.1024B}.  Each used further explored concepts abandoned in early \superfit{}, development --- a focus on key features in {\it GELATO}, and cross-correlation in the case of {\it SNID}, adding complementary approaches. All three codes were widely adopted in the community.

Recently, supernova fitting codes based on deep learning have arisen, notably Deep Learning for the Automated Spectral Classification and Their Hosts \cite[DASH;][]{2019ApJ...885...85M}.  It now exists in an updated web-based form called {\it Astrodash}\footnote{https://astrodash.scimma.org}.  Such deep learning approaches are promising for their speed, flexibility, and accuracy.  However, there are drawbacks, including understanding what the model is doing, difficulty in retraining models, and biases and errors that can creep into the training.  There is room for multiple approaches.

%Spectroscopic classification is central to the physical interpretation of supernovae (SNe) and other transients. In practice, classification can be challenging when a transient is faint, the observed wavelength range covers only a subset of diagnostic features, or the spectrum contains substantial host-galaxy contamination. The original \superfit{} addressed these challenges through a template-fitting approach that compares an observed spectrum with libraries of SN spectra spanning multiple types and phases, together with host-galaxy spectra. Redshift, reddening, and host-galaxy contributions are varied to identify the best-fitting combination by minimizing a $\chi^2$-based residual statistic \citep{Howell2005}. \newline

The original implementation of \superfit{} described by \citet{Howell2005} included 184 Type Ia spectra, 75 Type Ib/c spectra, and 47 Type II spectra, selected to cover a range of phases, wavelengths, and known subtypes.  Since then, the diversity of transient classes and the availability of observational spectra have increased substantially, and the comparison library has grown. Meanwhile, the astronomical community has shifted to Python.  A command-line driven, Python implementation of \superfit{}, called {\it Next Generation Superfit}, featuring an updated algorithm and performance improvements, has been available for several years \citep{2022TNSAN.191....1G}.  Here we characterize some of those improvements and introduce and upgraded version of \superfit{} with a web-based Graphical User Interface\footnote{https://superfit.space}.  

The updated software preserves the established template-fitting methodology while improving computational performance, expanding coverage of transient subclasses, and providing greater flexibility in fitting configurations. The web application makes \superfit{} accessible without local installation, allowing users to upload spectra, configure fitting parameters, and interactively examine and compare candidate matches. The code is open source and is available on GitHub\footnote{https://github.com/dahowell/WebSuperfit}.

\section{Methodology}\label{sec:method}
\subsection{Spectral Model}

The fitting procedure follows the original \superfit{} methodology described by \citet{Howell2005}, with some modifications. Each observed spectrum is modeled as a linear combination of a transient template and an optional host-galaxy template,

\begin{equation}
F_{\mathrm{model},i}
= a\,T_{\mathrm{SN}}(\lambda_i;z,A_V)
+ b\,T_{\mathrm{gal}}(\lambda_i;z),
\label{eq:model}
\end{equation}

where $T_{\mathrm{SN}}$ and $T_{\mathrm{gal}}$ are the transient and host-galaxy templates, $z$ is the redshift, $A_V$ is the reddening parameter, and $a$ and $b$ are the component scaling coefficients. Both templates are redshifted by $(1+z)$ and interpolated onto the observed wavelength grid. A wavelength-dependent extinction correction based on the \citet{1989ApJ...345..245C} law with $R_V=3.1$. 

For each combination of transient template, galaxy template, redshift, and reddening, the scaling coefficients $a$ and $b$ are determined analytically by least squares fitting. The agreement between the resulting model and the observed spectrum is then evaluated using

\begin{equation}
\chi^2 =
\sum_{i \in \mathcal{V}}
\frac{[F_{\mathrm{obs},i}-F_{\mathrm{model},i}]^2}
{\sigma_i^2},
\label{eq:chi2}
\end{equation}

where $\mathcal{V}$ contains the overlapping wavelength bins and $\sigma_i$ is the adopted flux uncertainty, either supplied with the spectrum or estimated from the data. Candidate fits are ranked using

\begin{equation}
Q = \frac{\chi^2}{(N-p)^2},
\label{eq:ranking}
\end{equation}

where $N$ is the number of overlapping wavelength bins and $p$ is the number of fitted component coefficients ($p=1$ for transient-only fits and $p=2$ when a galaxy component is included). The additional factor of $(N-p)$ favors fits with broader wavelength overlap, reducing the preference for templates that reproduce only a small portion of the observed spectrum. Because supernova spectra vary even within the same class, a good spectral match does not necessarily yield a statistically acceptable chi-square. Therefore,  we use ranking instead of assigning probabilities to the classifications.

In the command line version, users may weight each flux point by the inverse variance.  This can be provided as a third column in the input spectrum, can be specified as a constant, or it can be estimated from the data.   

\subsection{Redshift, Reddening, Phase, and Template Grids}

The redshift may be interated over a user-defined interval and step size, or fixed when a redshift is known. The reddening parameter is similarly evaluated over a user-defined grid. For each redshift--reddening combination, the selected transient and galaxy templates are transformed, interpolated onto the common wavelength grid, fitted, and ranked. The template search can be restricted to selected transient subtypes, selected galaxy types, and a desired range of spectral phases. Template phases are estimated from spectrum dates and dates of maximum brightness, when available, allowing templates to be filtered by a user-selected phase range.

\subsection{Template Library}

The updated \superfit{} template library contains 1008 spectra including Type Ia, Ib, II, and superluminous supernovae, tidal disruption events and fast blue optical transients. Table~\ref{tab:template_bank} summarizes the number of spectra available for each subtype and
their spectral epoch ranges. The library includes both observed spectra and synthetic models.
\section{Software Implementation and Web Interface}
\label{sec:superfit}
\subsection{Python Implementation}

We reorganized the fitting workflow into a Python package that handles preprocessing, template selection, spectral
fitting, and ranked output generation (Figure 1). The backend runs locally or is accessible online through the web application, supports fits with or without a host-galaxy component, and saves the JSON configuration and ranked results for reproducibility. The web application is implemented as a multi-page Dash (the Python web GUI software; not the deep learning supernova classifcation software of the same name) interface with Input and Output pages for uploading spectra,
configuring fits, monitoring progress, and inspecting results (Figure 2). Front end passes the
spectrum and selected settings to the Python backend, which performs the fitting calculation.

% FIGURE 1: web_workflow
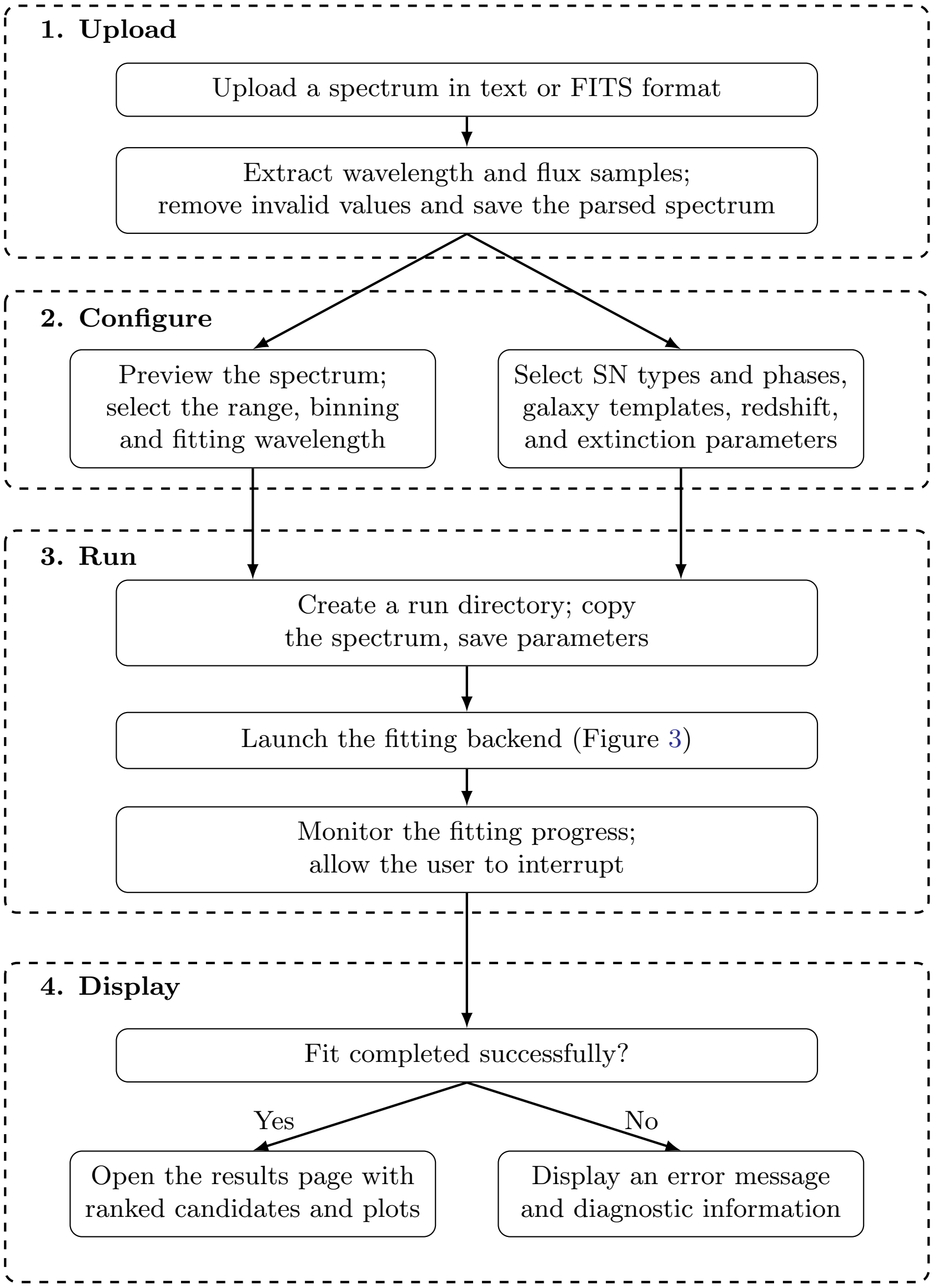
\begin{figure*}[!ht]
\centering
\begin{tikzpicture}[
    x=1cm, y=1cm,
    font=\small,
    box/.style={
        draw, rounded corners,
        align=center, text width=8cm,
        inner sep=5pt
    },
    branch/.style={box, text width=4cm},
    arrow/.style={->, >=latex, line width=0.8pt},
    stage/.style={
        draw, thick, rounded corners,
        fill=none
    },
    heading/.style={anchor=west, font=\small\bfseries}
]

\draw[stage, dashed] (-5.5,0) rectangle (5.5,-3);
\node[heading] at (-5.2,-0.3)
    {1. Upload};

\draw[stage, dashed] (-5.5,-3.4) rectangle (5.5,-5.75);
\node[heading] at (-5.2,-3.74)
    {2. Configure};

\draw[stage, dashed] (-5.5,-6.25) rectangle (5.5,-10.8);
\node[heading] at (-5.2,-6.55)
    {3. Run};

\draw[stage, dashed] (-5.5,-11.4) rectangle (5.5,-15.2);
\node[heading] at (-5.2,-11.7)
    {4. Display};

% Nodes
\node[box] (upload) at (0,-1)
    {Upload a spectrum in text or FITS format};

\node[box] (parse) at (0,-2.2)
    {Extract wavelength and flux samples;\\
     remove invalid values and save the parsed spectrum};

\node[branch] (preview) at (-2.55,-4.8)
    {Preview the spectrum;\\
     select the range, binning\\
     and fitting wavelength
     };

\node[branch] (settings) at (2.55,-4.8)
    {Select SN types and phases,\\
     galaxy templates, redshift,\\
     and extinction parameters};

\node[box] (prepare) at (0,-7.35)
    {Create a run directory; copy the spectrum, save parameters};

\node[box] (run) at (0,-8.75)
    {Launch the fitting backend (Figure~\ref{fig:fitting_workflow})};

\node[box] (monitor) at (0,-10.05)
    {Monitor the fitting progress;\\
     allow the user to interrupt};

\node[box] (success) at (0,-12.5)
    {Fit completed successfully?};

\node[branch] (results) at (-2.55,-14.15)
    {Open the results page with 
     ranked candidates and plots};

\node[branch] (error) at (2.55,-14.15)
    {Display an error message\\
     and diagnostic information};

% Connections
\draw[arrow] (upload.south) -- (parse.north);
\draw[arrow] (parse.south) -- (preview.north);
\draw[arrow] (parse.south) -- (settings.north);

\draw[arrow] (preview.south)
    -- (preview.south |- prepare.north);
\draw[arrow] (settings.south)
    -- (settings.south |- prepare.north);

\draw[arrow] (prepare.south) -- (run.north);
\draw[arrow] (run.south) -- (monitor.north);
\draw[arrow] (monitor.south) -- (success.north);

\draw[arrow] (success.south)
    -- node[pos=0.55, left, xshift=-15pt] {Yes}
    (results.north);

\draw[arrow] (success.south)
    -- node[pos=0.55, right, xshift=10pt] {No}
    (error.north);

\end{tikzpicture}
\caption{The four stages of the \superfit{} web workflow. The frontend handles spectrum upload, fit configuration, progress monitoring, and result display. The backend performs the fitting calculation summarized in Figure~\ref{fig:fitting_workflow}.}
\label{fig:web_workflow}
\end{figure*} \newpage

\subsection{Interface}

\begin{figure*}[!htt]
    \centering
    \begin{minipage}{0.49\textwidth}
        \centering
        \textbf{(a) Input}
    \end{minipage}%
    \begin{minipage}{0.49\textwidth}
        \centering
        \textbf{(b) Output}
    \end{minipage}
\includegraphics[width=\textwidth]{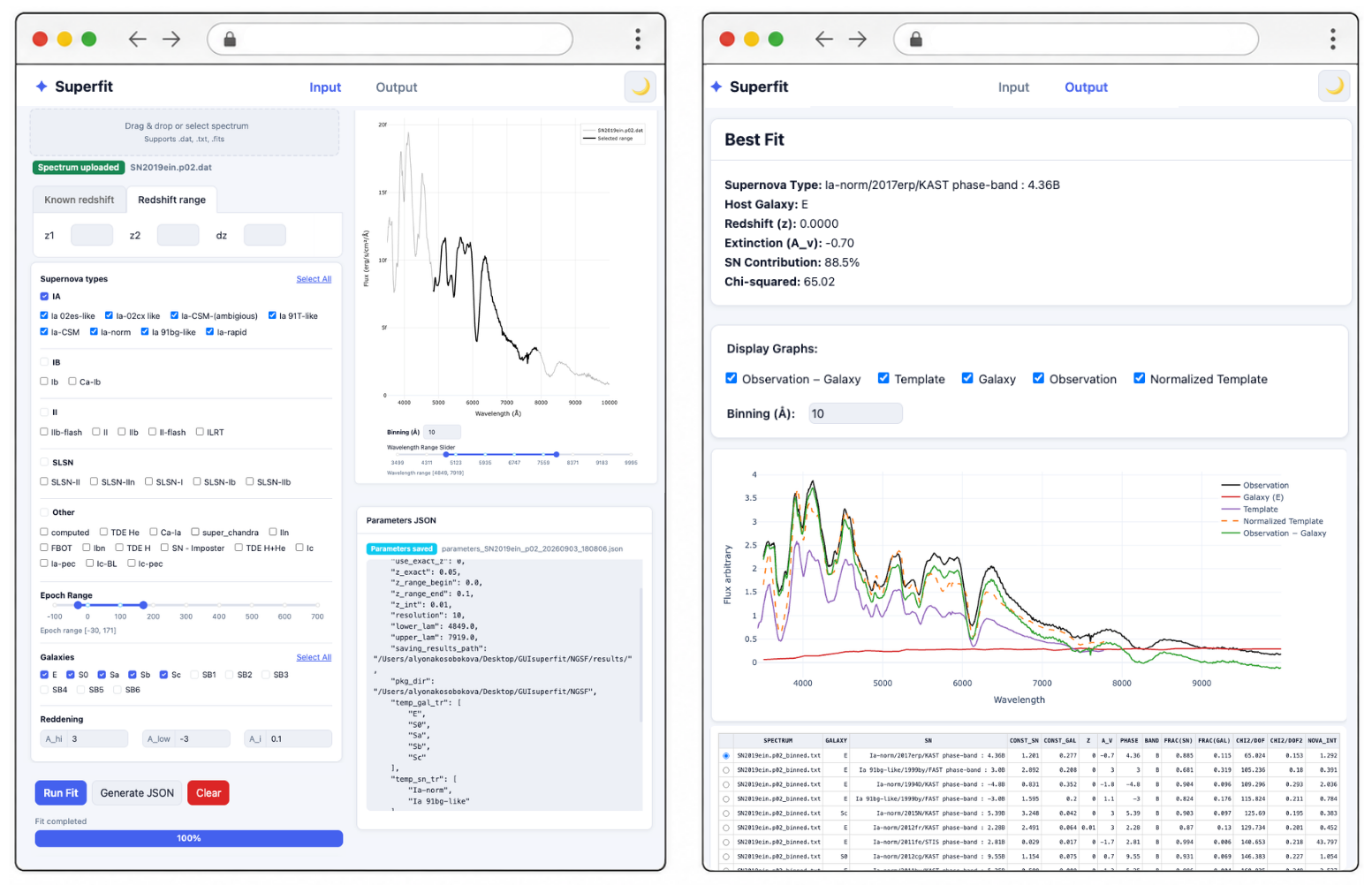}
\caption{The \superfit{} web interface:
(a) the Input page for spectrum preview and fit configuration;
(b) the Output page for ranked results and spectral inspection.}
\label{fig:web_pages}
\end{figure*}

The Input page (Figure~\ref{fig:web_pages}a) accepts ascii or FITS spectra and provides an interactive preview. Users select the fitting wavelength range and bin width, transient subtypes and template phases, and optional host-galaxy templates. The redshift can be fixed or searched over a specified range. Users can download the JSON configuration or launch a fit directly from the interface.  The spectrum is restricted to the requested wavelength interval and rebinned using a user-specified bin width, with a default of 10\,\AA{}. The rebinned flux is calculated as the median flux within each output wavelength bin. Before fitting, the observed flux is normalized by its median. 
%A fit is rejected if the requested bin width is smaller than the minimum native wavelength spacing or if the spectrum contains fewer than 35 estimated wavelength bins. The application prepares a separate run directory with the input spectrum and selected parameters. 
A progress bar displays the state, and the interface allows the user to stop a running fit.

The Output page (Figure~\ref{fig:web_pages}b) displays the ranked candidate table and selects the best fit result by default. Each result shows the transient and host galaxy templates, scaling coefficients, redshift, extinction, template phase (and band it is relative to), estimated component light fractions, and fit statistics. Users can select other candidates and inspect the observed spectrum, transient template, fitted transient and galaxy components, and the observation after subtraction of the fitted galaxy contribution. Display sampling can be adjusted without rerunning the fit.

 \newpage

\subsection{Backend Fitting}

The backend is organized around three main functions. The \texttt{all\_parameter\_space()} function iterates over a redshift--extinction grid, which contains all combinations of the user-specified redshift and extinction values, or their defaults. For each combination, it calls \texttt{core()} to evaluate the fits. The \texttt{sn\_hg\_arrays()} function applies the corresponding transformations to the transient and galaxy templates and interpolates them onto the observed wavelength grid. Extinction curves are precomputed for the transient templates and reused throughout the search. \newline

The \texttt{core()} function evaluates transient--galaxy template pairs using vectorized NumPy operations. For each redshift--extinction combination, the scaling coefficients (see $a$ and $b$ in Equation~\ref{eq:model}) are calculated using least squares,
and the resulting models are compared with the observed spectrum using Equation~\ref{eq:chi2}.  The method retains the 10 template pairs with the best fitting statistic $Q$, defined in Equation~\ref{eq:ranking}. Fits  with less than 50\% wavelength overlap are rejected.  After all combinations are evaluated, the best-scoring fit for each transient template is selected. The final results are ranked by increasing $Q$, with lower values indicating better fits. The backend workflow is summarised in Figure~\ref{fig:fitting_workflow}.

\begin{figure*}[!ht]
\centering
\begin{tikzpicture}[
    node distance=5mm,
    box/.style={
        rectangle, draw, rounded corners,
        align=center, text width=7cm,
        inner sep=4pt, font=\small
    },
    branch/.style={box, text width=4cm},
    decision/.style={
        diamond, draw, aspect=3,
        align=center, text width=3.4cm,
        inner sep=2pt, font=\small
    },
    stage/.style={
        rectangle, draw, thick, rounded corners,
        inner sep=4mm
    },
    stageLabel/.style={
        font=\small\bfseries,
        anchor=south west
    },
    arrow/.style={->, >=latex, line width=0.9pt},
]
% Stage 1: Preparation
\node[box] (templates)
    {Prepare observed spectrum and templates;\\
precompute extinction curves};

% Stage 2: Parameter search
\node[box, below=10mm of templates] (trial)
    {Iterate over redshift--extinction pairs $(z,A_V)$;\\
   for each call \texttt{core()} function };

\node[branch, below=of trial, xshift=-2.5cm] (sn)
    { Transient template apply\\
    extinction and redshift};

\node[branch, below=of trial, xshift=2.5cm] (gal)
    {Galaxy template \\
     apply redshift};

\node[box, below=24mm of trial] (grid)
    {Interpolate both components onto\\
     the observed wavelength grid};

% Stage 3: Fitting (core)

\node[box, below=of grid] (fit)
    {Calculate scaling coefficients $a$, $b$\\
    for all pairs using vectorized least squares
    (Equation~\ref{eq:model})};

\node[box, below=of fit] (score)
    {Evaluate fits using $\chi^2$
    (Equation~\ref{eq:chi2});\\
    retain the 10 lowest-scoring pairs by $Q$
    (Equation~\ref{eq:ranking})};

\node[decision, below=of score] (more)
    {More parameter\\
combinations?};

\coordinate (loopLeft) at (-5.7,0);
\coordinate (loopTop) at (loopLeft |- trial.west);
\coordinate (loopBottom) at (loopLeft |- more.west);

% Final results
\node[box, below=13mm of more] (rank) {
    Retain the best fit for each transient template;\\
    rank and save results
};

% Connections
\draw[arrow] (templates) -- (trial);
\draw[arrow] (trial.south) -- (sn);
\draw[arrow] (trial.south) -- (gal);
\draw[arrow] (sn.south) -- (grid);
\draw[arrow] (gal.south) -- (grid);
\draw[arrow] (grid) -- (fit);
\draw[arrow] (fit) -- (score);
\draw[arrow] (score) -- (more);

\draw[arrow] (more.west)
    -- node[above, font=\small] {Yes} (loopBottom)
    -- (loopTop) -- (trial.west);

\draw[arrow] (more.south)
    -- node[right, pos=0.25, font=\small] {No} (rank.north);

\end{tikzpicture}
\caption{\superfit{} backend flow.}
\label{fig:fitting_workflow}
\end{figure*}

\section{Examples of Classification}\label{sec:validation}

The following examples illustrate uses of \superfit{}, including host-galaxy decomposition, classification of objects excluded from the template library, and fitted with and without a supplied redshift. Examples are intended as a overall \superfit{} capabilities demonstration, and not a statistical assessment of classification accuracy.  No spectra shown here as examples were used in the comparison library.

\begin{deluxetable*}{llccll}[!ht]
\tablecaption{Examples of Superfit Classification
\label{tab:example_results}}
\tabletypesize{\scriptsize}
\tablewidth{0pt}
\tablehead{
\colhead{Name} &
\colhead{Best-fit Template} &
\colhead{Reference Phase (days)} &
\colhead{Best-template Phase (days)} &
\colhead{Reference Type} &
\colhead{Recovered Type}
}
\startdata
SN 2010kd & PTF12dam / LRIS & $-27.97$\ & $-21.0$ & SLSN-I & SLSN-I \\
SN 2020acat & 2008ax / FAST & $-13.77 \pm 0.87$ & -$14.85$ & IIb & IIb \\
SN 2011fe & 2011by / KAST & $-10.75$ & $-10.76$ & Ia & Ia \\
SN 2019yz & 2007gr / KAST & $-8.60$ & $-2.49$ & Ic & Ic \\
OGLE-2013-SN-123 & 2011by / KAST & $-0.07 \pm 0.86$ & $+5.25$ & Ia & Ia \\
\enddata
\tablecomments{
Best-template phases are reported as returned by \superfit{} and refer to the template objects phase.
}
\end{deluxetable*}

\subsection{SN 2010kd: SLSN Classification Without a Supplied Redshift}

\begin{figure}[H]
    \centering
    \includegraphics[width=\linewidth]{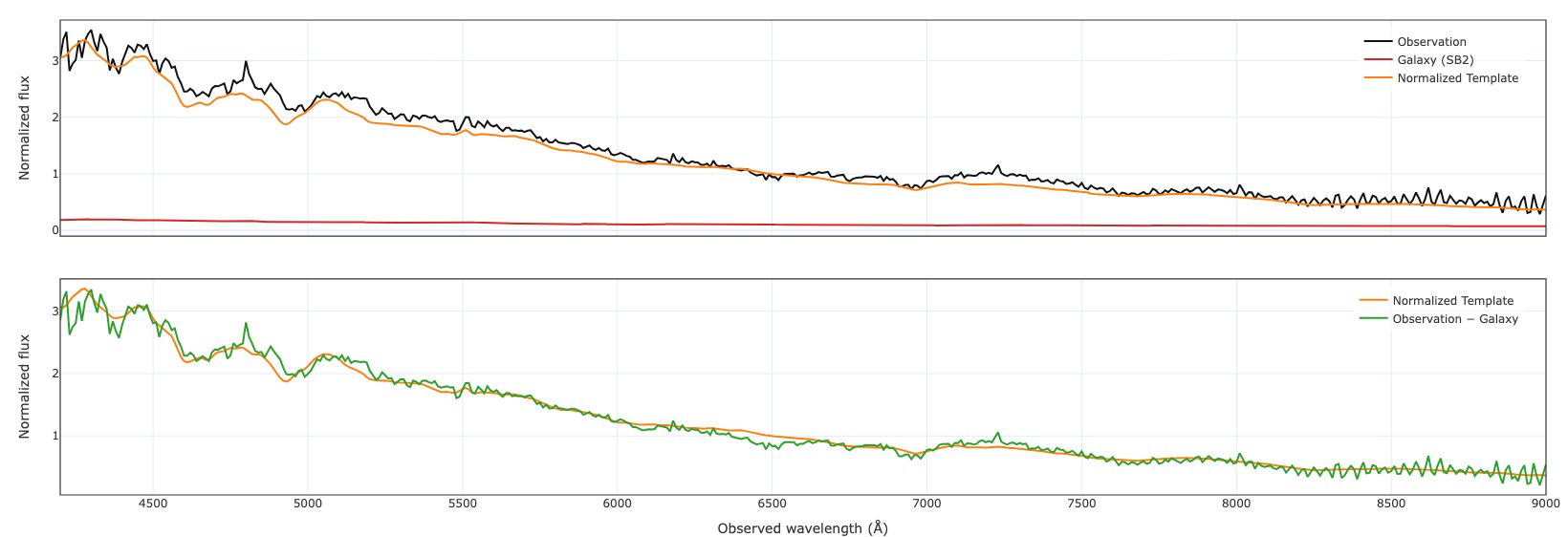}
    \caption{
    Best-fitting decomposition of SN 2010kd with redshift treated as a free parameter. The fit selects a PTF12dam template and recovers $z=0.100$.
    }
    \label{fig:SN_2010kd}
\end{figure}

SN 2010kd illustrates classification without a supplied redshift (Fig.\ref{fig:SN_2010kd}). \superfit{} recovers the SLSN-I classification and a redshift of $z=0.100$, close to the published value of $z=0.101$
\citep{Kumar2020}. The best-fitting template is PTF12dam at $-21.0$ days relative to its $B$-band maximum. The observation corresponds to $-27.97$ observer-frame days, or $-25.40$ rest-frame days, before the $B$-band maximum of SN 2010kd.

\subsection{SN 2020acat: Early Type IIb Classification}

\begin{figure}[H]
    \centering
    \includegraphics[width=\linewidth]{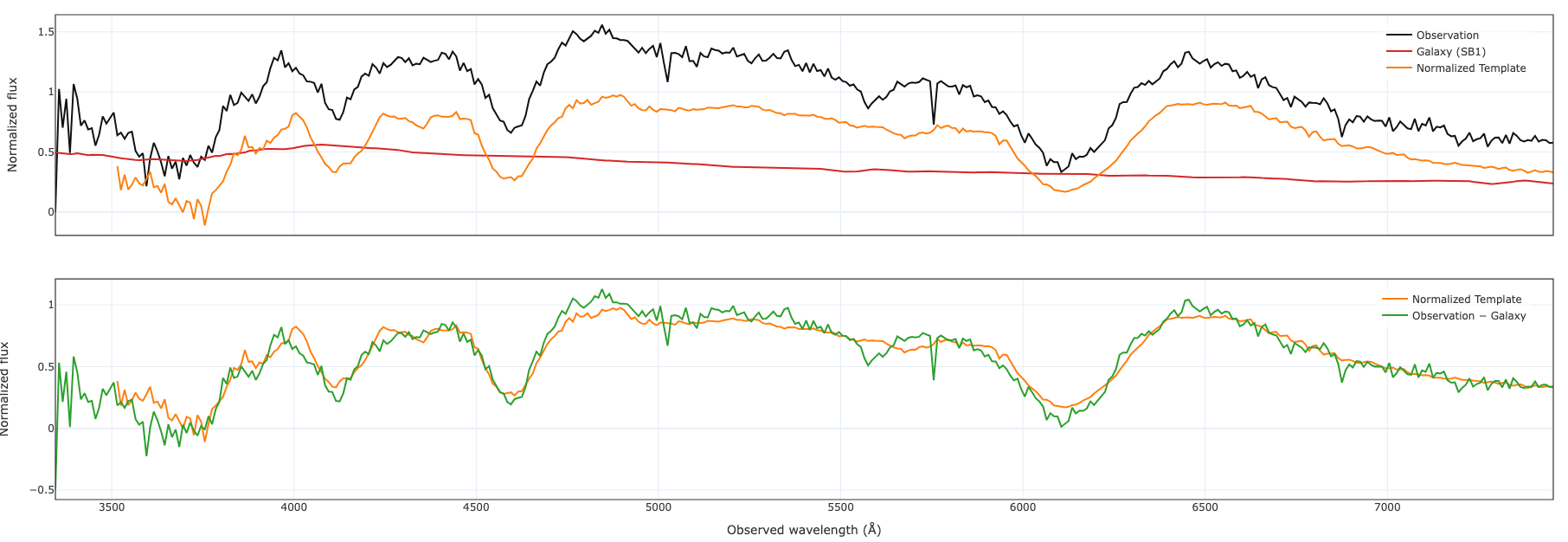}
    \caption{
    Best-fitting decomposition of the early spectrum of SN 2020acat
    }
    \label{fig:SN_2020acat}
\end{figure}

The spectrum of SN 2020acat corresponds to $-13.77 \pm 0.87$ rest-frame days relative to
$B$-band maximum, using the peak date reported by \citet{Medler2022}. \superfit{} recovers the Type IIb
classification, and selects SN 2008ax/FAST at $-14.85$ days relative to its $B$-band maximum (Fig.\ref{fig:SN_2020acat}).

\subsection{SN 2019yz: Type Ic Classification}

\begin{figure}[H]
    \centering
    \includegraphics[width=\linewidth]{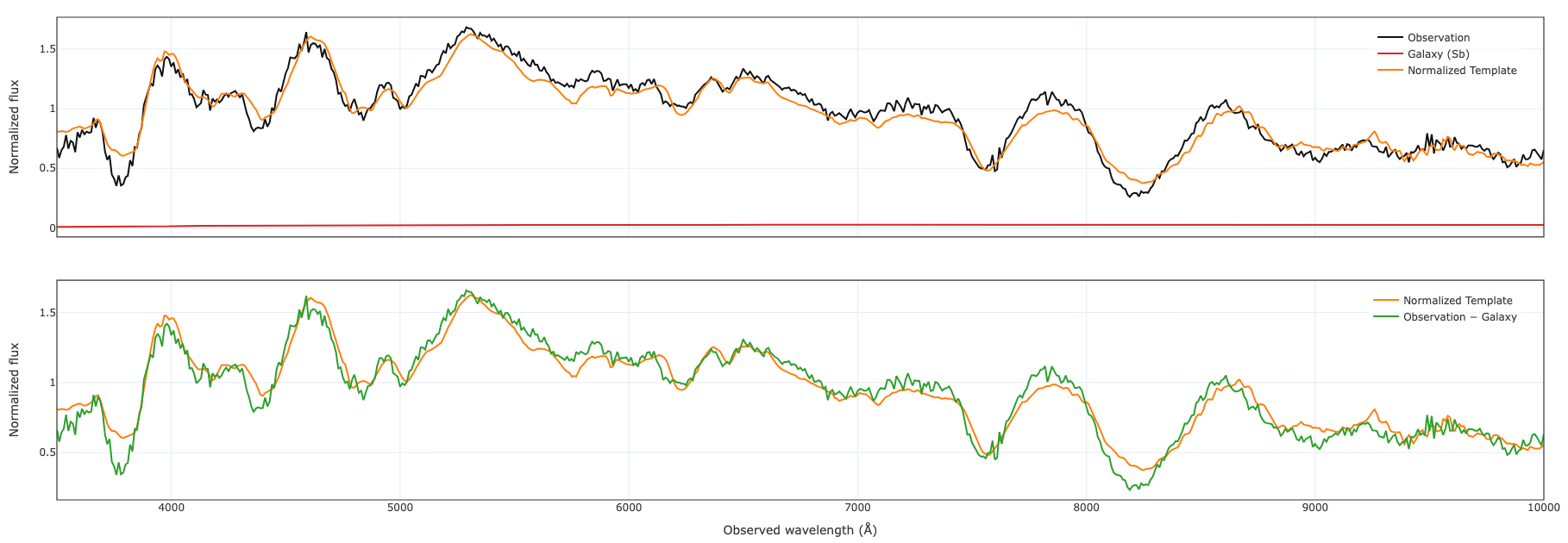}
    \caption{
    Best-fitting decomposition of SN 2019yz using a Type Ic SN 2007gr template and an Sb galaxy template. Redshift $z=0.0064$
    }
    \label{fig:SN_2019yz}
\end{figure}

The spectrum of SN 2019yz obtained on 2019 January 24 with a phase of $-8.6$ days relative to
$V$-band maximum \citep{Williamson2023}. \superfit{} recovers the Type Ic classification, selecting SN 2007gr/KAST at $-2.49$ days relative to its $B$-band maximum (Fig.\ref{fig:SN_2019yz}.

\subsection{SN 2011fe: Early Type Ia Classification}

\begin{figure}[H]
    \centering
    \includegraphics[width=\linewidth]{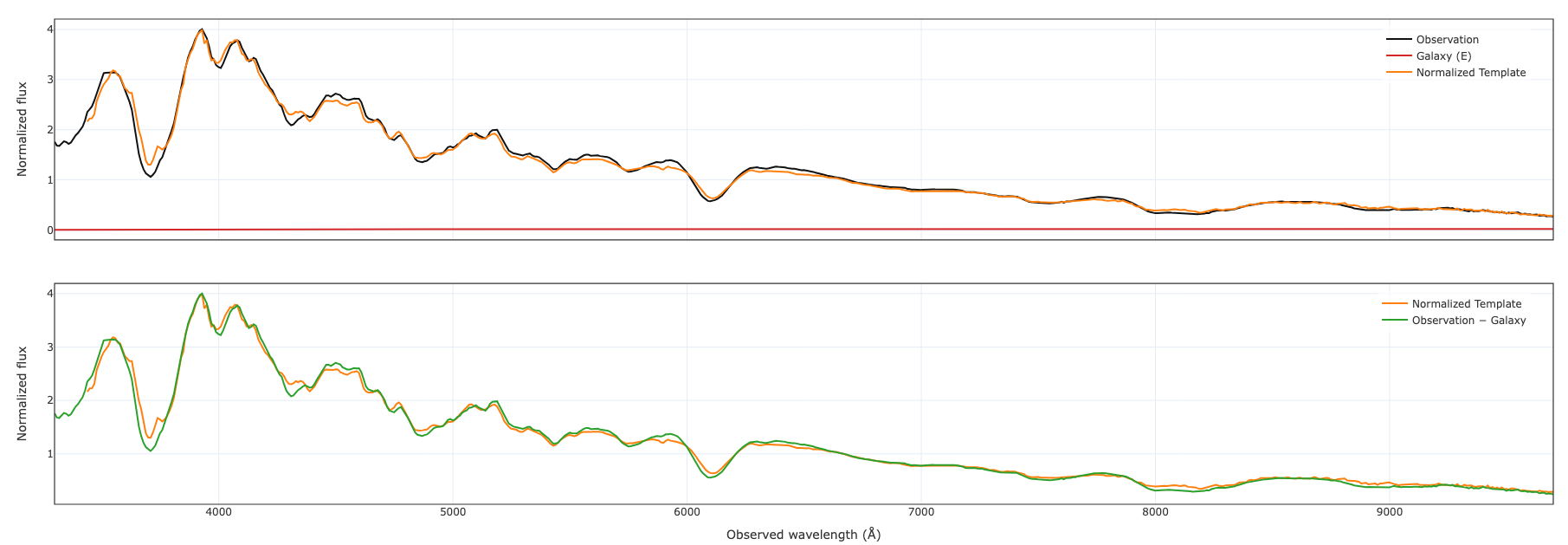}
    \caption{
    Best-fitting decomposition of an early spectrum of SN 2011fe, with all SN 2011fe templates excluded. SN 2011fe and SN 2011by are optical twins.
    }
    \label{fig:SN_2011fe}
\end{figure}

All spectra of SN 2011fe were excluded from the template library to test classification using spectra of other supernovae. \superfit{} recovers the Type Ia classification both with and without a supplied redshift of $z=0.0008$. The fit shown in Figure~\ref{fig:SN_2011fe} attributes 98.5\% of the fitted flux to the supernova and selects SN 2011by/KAST at $-10.76$ days relative to $B$-band maximum. SN 2011fe and SN 2011by are known optical twins, which explains their close spectral match \citep{2015MNRAS.446.2073G}.

\subsection{OGLE-2013-SN-123: Host-Galaxy Decomposition with Phase Uncertanty}

\begin{figure}[H]
    \centering
    \includegraphics[width=\linewidth]{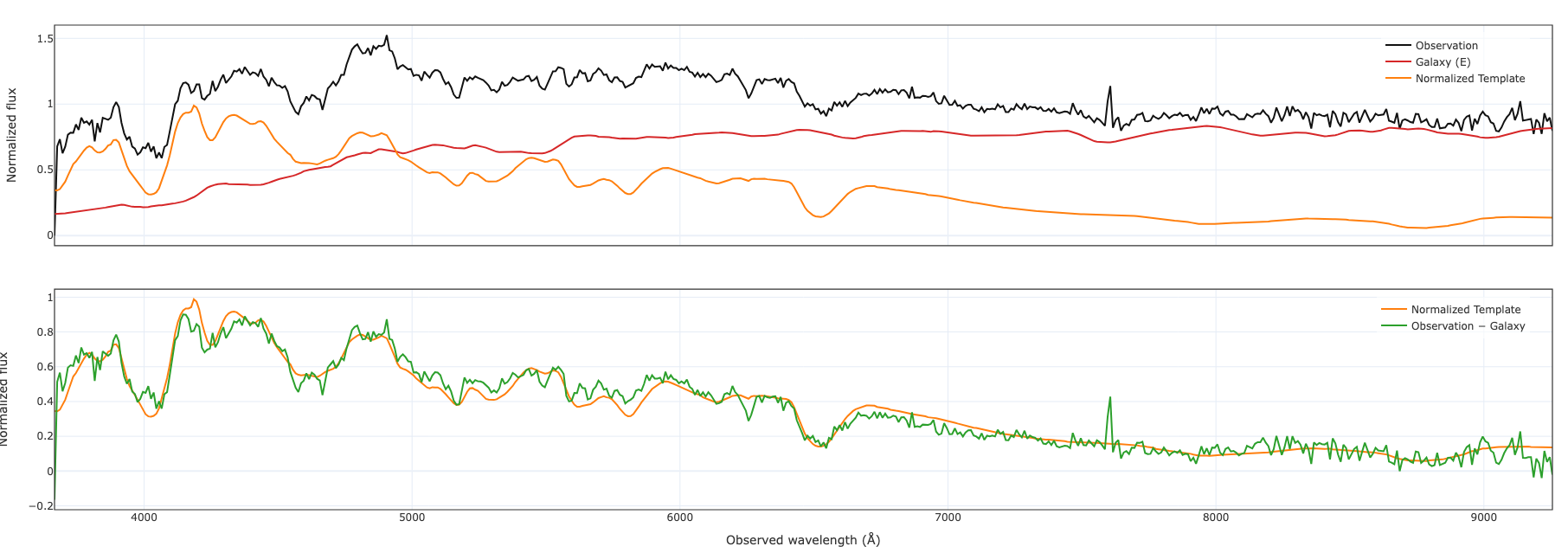}
    \caption{
    Decomposition of OGLE-2013-SN-123 at fixed $z=0.0614$.(a) Observed spectrum, fitted SN-plus-galaxy model, and fitted galaxy component. (b) Galaxy-subtracted observation compared with the fitted SN component. The fitted galaxy contributes 65.1\% of the model flux.
    }
    \label{fig:ogle_fit}
\end{figure}

For OGLE-2013-SN-123, 
despite a significant host-galaxy contribution of 65.1\%, \superfit{} recovers the Type Ia classification both with and without a supplied redshift of $z=0.0614$ (Fig. \ref{fig:ogle_fit}). For the fixed-redshift fit, the best-fitting template is
SN 2011by/KAST at $+5.25$ days relative to $B$-band maximum, compared with the photometric phase of $-0.07 \pm 0.86$ days. This example demonstrates Superfit's ability to reliably classify supernovae with significant host-galaxy contamination while highlighting the limitations of less accurate phase estimation with strong host contamination.

\section{Conclusions}\label{sec:conclusions}
The updated \superfit{} combines a Python implementation, an expanded template library, and an interactive web interface. Users can fit spectra across a range of parameters and inspect the results without local installation, or run the software locally without trained machine-learning models. 

While deep learning classification techniques are gaining prevalence, \superfit{} provides a complementary approach.  Users have fine-tuned control over every aspect of the fit, and can mathematically understand what the code is doing and the computations involved.  They can inspect the individual components of the fits to gain physical insight or ascertain the accuracy of the results.  The model also doesn't require retraining, and is not as susceptible to bias from a lopsided training set.  
%Classification depends on the coverage of the template library. Underrepresented types and phases may lack suitable matches, while spectral similarities between different types and epochs can produce ambiguous results. The highest-ranked template therefore requires inspection, and its phase should not be treated as a precise age measurement. Differences in phase references and observer- versus rest-frame conventions must also be accounted for when comparing template phases with independent estimates.
The web-based version of \superfit{} has the advantage that the user is not required to install any software, and improvements can be rolled out to all users instantaneously.

In the era of big data and automated data flows, \superfit{} can help to immediately classify supernovae without a human in the loop, speeding up response times for transient study.  The Python version has already been incorporated into archives such as WISERep \citep{YaronGalYam2012}.  Future integration into Target and Observation Managers (TOMs) via the TOM Toolkit \citep{2024ASPC..535....3S} is also planned.  Additional improvements will focus on expanding coverage of underrepresented types and phases, standardizing phase metadata, and refining the fitting and classification methods.

\begin{deluxetable*}{|llcc|}[!ht]
\tablecaption{Superfit Transient Spectral Template Bank
\label{tab:template_bank}}
\tabletypesize{\scriptsize}
\tablewidth{0pt}
\tablehead{
\multicolumn{1}{|c}{Type} &
\colhead{Subtype} &
\colhead{Number of Spectra} &
\multicolumn{1}{c|}{Epoch Range (days)}
}
\startdata
\textbf{Ia} & Ia 02es-like & 16 & $-14.12$ to $+90.92$ \\
 & Ia-02cx like & 23 & $-8.59$ to $+54.28$ \\
 & Ia-CSM-(ambigious) & 9 & $+2.00$ to $+43.00$ \\
 & Ia 91T-like & 19 & $-13.00$ to $+257.00$ \\
 & Ia-CSM & 13 & $-1.00$ to $+88.00$ \\
 & Ia-norm & 89 & $-16.99$ to $+278.70$ \\
 & Ia 91bg-like & 33 & $-6.87$ to $+183.65$ \\
 & Ia-rapid & 4 & $-3.77$ to $+22.53$ \\
 & Ia 99aa-like & 14 & $-14.71$ to $+285.00$ \\
\textbf{} & & \textbf{220} & {} \\
\hline
\textbf{Ib} & Ib & 63 & $-31.00$ to $+1017.34$ \\
 & Ca-Ib & 21 & $-1.65$ to $+126.23$ \\
\textbf{} & & \textbf{84} & \\
\hline
\textbf{II} & IIb-flash & 3 & $-7.00$ to $-1.06$ \\
 & II & 127 & $-76.00$ to $+356.26$ \\
 & IIb & 89 & $-18.00$ to $+398.61$ \\
 & II-flash & 22 & $-6.87$ to $+3.19$ \\
 & ILRT & 13 & $-20.77$ to $-3.63$ \\
\textbf{} & & \textbf{254} & \\
\hline
\textbf{SLSN} & SLSN-II & 15 & $-13.00$ to $+427.06$ \\
 & SLSN-IIn & 20 & $-31.39$ to $+142.39$ \\
 & SLSN-I & 96 & $-28.66$ to $+438.34$ \\
 & SLSN-Ib & 8 & $-32.00$ to $+263.00$ \\
 & SLSN-IIb & 2 & $-29.81$ to $+22.78$ \\
\textbf{} & & \textbf{141} &  \\
\hline
\textbf{Other}  & TDE He & 2 & $-25.67$\\
 & Ca-Ia & 9 & $+0.17$ to $+0.16$ \\
 & IIn & 45 & $-2.00$ to $+845.34$ \\
 & Ibn & 35 & $-18.68$ to $+45.82$ \\
 & TDE H+He & 24 & $-12.00$ to $+199.30$ \\
 & Ic & 47 & $-9.00$ to $+172.00$ \\
 & Ia-pec & 27 & $-5.00$ to $+146.17$ \\
 & Ic-BL & 50 & $-5.60$ to $+201.00$ \\
 & Ic-pec & 5 & $-0.53$ to $+11.00$ \\
 & FBOT & 3 & $+5.6$ to $+24.5$ \\
 & super\_chandra & 37 & $-18.46$ to $+114.00$ \\
 & TDE H & 10 & \nodata \\
 & SN - Imposter & 6 & \nodata \\
 & computed (synthetic) & 9 & \nodata \\
\textbf{} & & \textbf{309}  &  \\
\hline
\textbf{Total} & & \textbf{1008} &  \\
\enddata
\tablecomments{Supernova subtypes with template spectra in \superfit , with the number of templates and phase ranges listed.}
\end{deluxetable*}

\begin{acknowledgments}
AGY’s research is supported by ISF, IMOS and BSF grants, as well as the André Deloro Institute for Space and Optics Research, the Center for Experimental Physics, a WIS-MIT Sagol grant, the Norman E Alexander Family M Foundation ULTRASAT Data Center Fund, and Magnus Konow in honor of his mother Olga Konow Rappaport;  AGY is the incumbent of the The Arlyn Imberman Professorial Chair.

DAH is supported by Las Cumbres Observatory.
\end{acknowledgments}

\begin{contribution}

\end{contribution}

\software{ Superfit, Astropy, Dash, Plotly, SciPy, pandas, Matplotlib, PyAstronomy, extinction}

\bibliography{superfit_refs.bib}
%\bibliography{references.bib}
\bibliographystyle{aasjournal}

\end{document}